\documentclass[twocolumn,nofootinbib]{revtex4-1}

\usepackage{mathrsfs}
\usepackage{amssymb}
\usepackage{amsmath}
\usepackage{bm}
\usepackage{graphicx}
\usepackage[usenames,dvipsnames]{color}
\usepackage[colorlinks=true,citecolor=blue,linkcolor=magenta]{hyperref}
\usepackage{placeins}
\usepackage{float}

\newcommand{\ket}[1]{\vert{ #1 }\rangle}

\def\add{\texttt{\,+\,}}

\newcounter{dummy}

\begin{document}

\title{Performance of a measurement-driven `adiabatic-like' quantum 3-SAT solver}

\author{Simon C. Benjamin}

\affiliation{Department of Materials, University of Oxford, Parks Road, Oxford OX1 3PH, United Kingdom}

\begin{abstract}
I describe one quantum approach to solving 3-satisfiability (3-SAT), the well known problem in computer science. The approach is based on repeatedly measuring the truth value of the clauses forming the 3-SAT proposition using a non-orthogonal basis. If the basis slowly evolves then there is a strong analogy to adiabatic quantum computing, although the approach is entirely circuit-based. To solve a 3-SAT problem of $n$ variables requires a quantum register of $n$ qubits, or more precisely {\it rebits} i.e. qubits whose phase need only be real. For cases of up to $n=26$ qubits numerical simulations indicate that the algorithm runs fast, outperforming Grover's algorithm and having a scaling with $n$ that is superior to the best reported classical algorithms.  There are indications that the approach has an inherent robustness versus imperfections in the elementary operations.
\end{abstract}

\maketitle

\section{Scope and nature of this document}
This document contains my notes on a series of numerical experiments I have performed to assess a quantum approach to solving classical 3-SAT problems. I use measurements to drive a register of qubits toward a target state satisfying a condition related to the 3-SAT (actually an instance of Bravyi's quantum 3-SAT~\cite{Bravyi06}). The target state can be slowly evolved from a trivial initial target to a final state that reveals the 3-SAT solution; this is analogous to adiabatic quantum computing~\cite{Farhi2000} but using only circuit-like operations. 

The numerical results I present are unverified -- they have not been independently reproduced, therefore either outright code errors or more subtle numerical inaccuracies may be occurring. I have not previously studied 3-SAT problems, therefore the instances that I generate may be the `wrong' choice and may include too many cases that are easy classically -- moreover I do not compare the performance of the quantum approach with a classical solver, I merely cite the reported performance bounds for such solvers. 

 The purpose of this document is to serve as a basis for discussion. I try to introduce all the terms/ideas I use without presuming any specific expertise from the reader, beyond undergraduate quantum physics and a basic understanding of algorithms.
For experienced quantum researchers, the short version is given in the box.

\section{Satisfiability}
Boolean satisfiability problems involve a `proposition' $P(x_1,x_2,...x_n)=\ ${\rm\small TRUE} where the $n$ boolean variables $x_i$ may each take the value {\rm\small TRUE} or {\rm\small FALSE}. Typically the challenge is to determine whether any choice of values for the variables $x_i$ can satisfy the formula.

In the particular case called 3-satisfiability (3-SAT) the expression $P(x_1,x_2,...x_n)$ is formed by AND'ing together a number of {\it clauses}, and each clause involves three variables OR'ed together. \\
{\bf Example:} $(x_1 \lor \lnot x_2 \lor x_3) \land (\lnot x_1 \lor x_4 \lor x_5) \land ... = {\rm TRUE}.$\\

Here the symbol  $\land$ denotes the logical AND, $\lor$ denotes the logical OR, while  $\lnot$ denotes the logical NOT which is also referred to as {\em negation}. Table~\ref{littleLogicTable} summarises these standard elements. Note that {\it verifying} a candidate solution merely involves  checking each independent clause is {\small TRUE}; if-and-only-if they all are, the 3-SAT is satisfied.

\smallskip

\vspace{4pt}
\setlength{\fboxsep}{10pt}
 \noindent\fbox{\parbox[c][14.1cm][c]{.9\linewidth}{
\noindent {\bf Overview of the paper}
\smallskip

I looked at an algorithm (actually a few variants of an algorithm) that seeks to solve 3-SAT problems by applying a series of post-selective measurements to a register of qubits. Each boolean of the 3-SAT corresponds to a unique qubit, and each quantum measurement involves checking that a clause of the 3-SAT is satisfied. The extra trick is that the single-qubit states representing {\small TRUE} and {\small FALSE} are not in general orthogonal. 

The process of performing a general clause check is shown in the box on page~\pageref{ccBox}. There is a unique state that must pass all the clause checks (Eqn.~\ref{fullTheta}) if the 3-SAT has a unique solution (and a corresponding subspace if multiple solutions, see Appendix). 

I've looked at a variant where the basis of {\small TRUE}/{\small FALSE} slowly evolves until becoming orthogonal, this is somewhat analogous to adiabatic quantum computing -- see the flow diagram Fig.~\ref{fig:alg_flow} and data in Figs.~\ref{fig:simpleLinearPart1} \& \ref{fig:simpleLinearPart2}. I've also looked at a rather simpler variant where one creates (`sculpts') the satisfying state for a fixed basis, before measuring it for partial information on the 3-SAT solution -- see flow diagram Fig.~\ref{fig:alg_sculpt} and Figs.~\ref{fig:timeTo999} \& \ref{fig:prefactors}. This latter approach is the one I've studied more fully using numerical experiments. I've also looked at a hybrid, see Figs.~\ref{fig:hybrid} and \ref{fig:2phaseData}.

Some of these approaches outperform both Grover's and leading conventional algorithms, {\it over the very limited range I can simulate} and with the caveat that I can't specify a fixed time within which the solution will certainly be found.

I have not properly investigated error tolerance, but I have tried adding 1\% or 2\% random over/under rotation to the simulation, and this seems well tolerated, see Fig.~\ref{noiseGraphs}.
}}

\begin{table}[!h]
\centering
\begin{tabular}{|cc|c|c|c|}
\hline
$\ \ x\ \ $ & $\ \ y\ \ $ & $\ \ \lnot x\ \ $  & $\ \ x \land y\ \ $ & $\ \ x \lor y\ \ $ \\
\hline
F & F & T & F & F \\
F & T & T & F & T \\
T & F & F & F & T \\
T & T & F & T & T\\
\hline
\end{tabular}
\caption{Table defining the AND, OR and NOT.}
\label{littleLogicTable}
\end{table}

\section{3-SAT Problems Considered Here}

In the analysis described here, I generate 3-SAT instances randomly but in such a way that

\begin{enumerate}
  \item \label{unique} All clauses are distinct from one another.
  \item \label{TandF} Every variable $x_i$ occurs at least once in the 3-SAT proposition in positive form $x_i$, and at least once in negated form $\lnot x_i$.
  \item \label{diffVar} All clauses  involve three different variables, i.e. neither $(x_1\lor x_1 \lor x_2)$ nor $(x_1\lor \lnot x_1 \lor x_2)$ would be generated.
\item \label{5times} Given $n$ variables I generate $N_c={\rm round}(R\ n)$ clauses, with $R=4.267$.
\end{enumerate}

Having generated a 3-SAT formula in this way I then determine the number of solutions, $N_S$. This is a trivial task classically for the problem sizes I consider. Generally I will be wish to have 3-SAT problems with a specific $N_S$ for testing the quantum algorithm; if the randomly generated 3-SAT has some other $N_S$ I discard it and generate a fresh formula. For the majority of data I present here I have chosen $N_S=1$, i.e. problems with a `unique satisfying assignment' (USA), however I do also consider cases with $N_S>1$. While it is quite rare to have exactly one solution, so that these 3-SAT instances are special, I choose this to reduce variably in the challenge of solving them. Note that criterion  (\ref{unique}) is simply to ensure proper counting of the number of clauses -- if two clauses were identical, one could be dropped without changing the 3-SAT problem. Similarly (\ref{TandF}) is imposed because otherwise the effective number of variables is reduced: for example if a variable only occurs in positive form $x_i$ without $\lnot x_i$ appearing anywhere in the formula then trivially one chooses $x_i$={\rm\small TRUE} and all clauses involving $x_i$ become ${\rm\small TRUE}$. Meanwhile (\ref{diffVar}) is merely for simplicity -- in principle recurrence of a variable within a clause is legitimate for 3-SAT but I do not consider such cases (although there is no apparent difficulty with such a generalisation). Point (\ref{5times}) is to maximise the chances that a randomly generated set of clauses will have a small but non-zero number of solutions $N_S$. I rely on the results of previous SAT studies which have found that there is a key threshold in the ratio $R$ between the number of clauses and the number of variables -- instances below the threshold are likely to be satisfied with multiple solutions, and instances above it are unlikely to be satisfiable. This threshold is still somewhat debated, and only becomes sharp as $n\rightarrow\infty$, but the value $4.267$ has been suggested~\cite{nature4p267}. 

\section{Classical 3-SAT Solvers}
The task of efficiently solving  3-SAT instances is a long studied challenge in computer science. It is believed that any algorithm will require a running time which is exponential in $n$~\cite{wikiETH}. Clearly the problem can be solved by a brute force search over all $2^n$ variable assignments. Then each failed test of a potential assignment prior to the solution will require some portion of the $N_c$ clauses to be checked, so that $t=N_c2^n$ is a trivial upper bound to the number of clause checks required. Throughout this paper I take the {\bf expected number of clause checks} to be the metric for the running time of an algorithm; this is fair only if the time required for a clause check is independent of the size of the problem, of course.

The na\"ive classical bound can be dramatically improved upon. A series of publications have succeeded in reducing the base $K$ in the the expression for the expected running time, $t\propto K^n$. The current record holder appears to be an algorithm due to Paturi {\it et al}~\cite{PaturiImproved2005} which boasts  $K=2^{2\ln 2-1}\approx1.307$.

\section{A Quantum 3-SAT Solver}

A simple application of Grover's search algorithm would yield a quadratic improvement over the brute force classical search, from $t\propto 2^n$ to $t\propto (\sqrt 2)^n$ i.e. $K=1.41$, but this is in fact inferior to the best classical solutions as noted above. It has been suggested~\cite{Ambainis} that one of the high performing classical algorithms can be accelerated by replacing the random search with a Grover-type coherent evolution, leading to a hybrid algorithm with running time $\propto (1.153)^n$.

Here I describe a different quantum approach based on repeatedly making projective measurements, each corresponding to evaluating the truth value of a generalised clause. I begin by assigning a qubit to represent each boolean variable $x_i$ in the problem. The general state of an isolated pure qubit is of course $\ket{\Psi}=\alpha\ket{0}+\beta\ket{1}$, where $\alpha$ and $\beta$ can be complex, but in this analysis I find it suffices to consider qubits on the real great circle of the Bloch sphere, i.e. where $\alpha$ and $\beta$ are real numbers. Qubits of this kind are sometimes called rebits. In the following I choose my basis so that, if I am using an orthogonal basis for representing truth values, then the state $\ket{1}$ will be identified with boolean {\rm\small TRUE}, and $\ket{0}$ will be identified with {\rm\small FALSE}. (Obviously, rotating all states and bases in what follows will yield the same algorithm with, say, $\ket{+}$ and $\ket{-}$ as the truth values, if that is preferable). In general I consider cases where {\rm\small TRUE} and  {\rm\small FALSE} are not orthogonal.

I begin by describing a na\"ive all-orthogonal approach to establish the framework. Consider a operation called ``Q-NOR$_Z$'', which involves 4 qubits, $a$, $b$, $c$, $d$. The operation simply applies a flip to qubit $d$ if, and only if, $a$, $b$ and $c$ are all in the $\ket{0}$ state; it does nothing otherwise. By ` flip' I refer to the Pauli $x$-operator. Therefore this four-qubit operation is merely a control-control-control-not gate acting when the controls are zeros.

\begin{table}[!h]
 \caption { Q-NOR$_Z$ operation}
 \centering
\begin{tabular}{|ccc|c|}
\hline
$\ \ a\ \ $ & $\ \ b\ \ $ & $\ \  c\ \ $  & \ \ action on $d$  \\
\hline
$\ket{0}$ & $\ket{0}$ & $\ket{0}$ &\  flip: $\ket{0}\leftrightarrow\ket{1}$\  \\
$\ket{0}$ & $\ket{0}$ & $\ket{1}$ & none  \\
$\ket{0}$ & $\ket{1}$ & $\ket{0}$ & none  \\
$\vdots$ & $\vdots$ & $\vdots$ & \ \vdots \\
$\ket{1}$ & $\ket{1}$ & $\ket{1}$ & none  \\
\hline
\end{tabular}
\label{QnorZ}
\end{table}
\smallskip

If I prepare $d$ in state $\ket{1}$ and apply the Q-NOR$_Z$ operation, and then measure $d$ in the $z$-basis (i.e. the basis with outcomes $\ket{0}$ and $\ket{1}$), then the outcome tells me the truth value of the $a\lor b\lor c$ clause: if-and-only-if $a$, $b$ and $c$ are all {\rm\small FALSE} then $d$ will be measured as $\ket{0}$, denoting {\rm\small FALSE}. If I wanted to test the clause $a\lor \lnot b\lor c$ then I would simply invert $b$ before this process (i.e. apply a $x-$gate to $b$), and again afterwards. What is the effect of this kind of process when the qubits $a$, $b$ and $c$ are in some general state? If we {\it fail} the clause check, by measuring $\ket{0}$ on the ancilla, then we project $a$, $b$ and $c$ onto state $\ket{000}$ (where I now write all three qubit states within a single ket for compactness). Conversely if we {\it pass} the check by measuring the ancilla in state $\ket{1}$, then we have projected out (removed) any component  $\ket{000}$ that had been present in the state of $a$, $b$, $c$.

Now consider the effect of a series such clause checks, one for each clause in our 3-SAT problem. Note that all such clause checks will {\it commute} with one another, because they are simply projections in the $z$-basis. Consequently if all clause checks are successfully passed then subsequently measuring the qubits in the $z$-basis will produce a satisfying assignment of truth values for the 3-SAT proposition.

Suppose that we initialise our qubits such that each is in the `plus state' $\ket{+}\equiv(\ket{0}\add\ket{1})/\sqrt{2}$, i.e. an equal superposition of our orthogonal {\rm\small TRUE} and {\rm\small FALSE} states. Then we perform each clause check, one after another, until {\it either} we see a `failed' outcome on the ancilla, {\it or} else we have passed one clause check for each clause in our 3-SAT problem. If indeed we succeed then we can directly measure the measure the qubits one at a time to obtain a satisfying solution; what is the probability of such a success? To answer this note that our initial state is simply the equal superposition of all states in the $z$-basis:
\[
2^{-\frac{n}{2}}\big(\ket{0\dots 00} \add \ket{0\dots 01} \add \dots \add \ket{1\dots 11} \big).
\]
If we are dealing with a 3-SAT that has a single solution, then all but one of these terms will be projected out by the succession of clause checks, and so the probability of success is simply the amplitude-squared on the correct term, $2^{-n}$. Therefore if we seek to solve our 3-SAT in this way, we expect to have to restart $\sim2^n$ times before finding a solution, equivalent to the worst case of the brute force classical approach.

Now I describe a simple generalisation of this setup: a quantum algorithm that is again based on repeated measurement of the truth value of clauses, but within a basis structure such that {\rm\small TRUE} and {\rm\small FALSE} are not generally orthogonal.  As a segue to that general case, first suppose that the basic multi-qubit operation that our computer can directly perform is not in fact ``Q-NOR$_Z$'' but rather ``Q-NOR$_X$'' as shown in the following table, which uses $\ket{-}\equiv(\ket{0}-\ket{1})/\sqrt{2}$.
\begin{table}[!h]
 \caption { Q-NOR$_X$ operation}
 \centering
\begin{tabular}{|ccc|c|}
\hline
$\ \ a\ \ $ & $\ \ b\ \ $ & $\ \  c\ \ $  & \ \ action on $d$  \\
\hline
$\ket{-}$ & $\ket{-}$ & $\ket{-}$ &\  flip: $\ket{0}\leftrightarrow\ket{1}$\  \\
$\ket{-}\ $ & $\ket{-}$ & $\ket{+}$ & none  \\
$\ket{-}\ $ & $\ket{+}$ & $\ket{-}$ & none  \\
$\vdots$ & $\vdots$ & $\vdots$ & \ \vdots \\
$\ket{+}\ $ & $\ket{+}$ & $\ket{+}$ & none  \\
\hline
\end{tabular}
\label{QnorX}
\end{table}
  
 \smallskip
  
Well we can simply recover the former clause check process by individually rotating the three control qubits appropriately, both before and after the application of ``Q-NOR$_X$''. We can use a $Y(\frac{\pi}{2})$ gate, where
\[
Y(\theta)=\left(\begin{array}{cc}\cos\frac{\theta}{2} & \sin\frac{\theta}{2}  \\ -\sin\frac{\theta}{2} & \cos\frac{\theta}{2} \end{array}\right)\ \ {\rm so}\ \ \ 
Y\left(\frac{\pi}{2}\right)=\frac{1}{\sqrt 2}\left(\begin{array}{cc}1 & 1  \\ -1 & \ 1 \end{array}\right).
\]
The $Y(\frac{\pi}{2})$ rotation maps state $\ket{0}$ to $\ket{-}$ and maps state $\ket{1}$ to $\ket{+}$. Therefore applying this operation, separately and independently, to the three qubits $a$, $b$ and $c$ will map from any of the input rows in Table (\ref{QnorZ}) to the corresponding row in Table (\ref{QnorX}). As before we prepare ancilla $d$ in state $\ket{1}$ prior to applying the Q-NOR$_X$ operation, then subsequently measure $d$ in the $z-$basis and say that the clause check is `passed' if-and-only-if the measured state is $\ket{1}$. Finally we  apply the inverse rotation $Y(-\frac{\pi}{2})$ to each of the qubits $a$, $b$, and $c$; this returns them to their original basis. 

Recall that in the na\"ive protocol if we have a clause in which one, or more, of the variables appears in negated form then we should invert the corresponding qubit(s) $\ket{0}\leftrightarrow\ket{1}$ before, and then after, the operation. But now that we are in any case applying rotations to our qubits, we notice that we can simply apply $Y(-\frac{\pi}{2})$ prior to the Q-NAND$_Z$ (rather than $Y(\frac{\pi}{2})$) to any qubit whose corresponding boolean appears in negated form in the clause. This maps $\ket{0}$ to $\ket{+}$ and correspondingly $\ket{1}$ to $(-1)\ket{-}$, where we note an extra phase of $-1$ in the latter case.  After the ancilla measurement we of course apply $Y(\frac{\pi}{2})$ to reverse the former rotation, and reabsorb the phase. 

Thus our earlier protocol for clause evaluation is equivalent to the following steps, if we choose $\theta=\frac{\pi}{2}$. I refer to this process as a {\it clause check} operation.
\vspace{5pt}
\setlength{\fboxsep}{12pt}
 \noindent\fbox{\parbox[c][5.5cm][c]{.9\linewidth}{
 \refstepcounter{dummy}
\noindent {\bf Clause check operation:}\label{ccBox}

\vspace{10pt}
1. Select a clause; let the three boolean variables to which it refers be $x_i$, $x_j$, $x_k$, and denote the corresponding qubits as $Q_i$, $Q_j$ and $Q_k$.

\vspace{6pt}
2. Rotate qubit $Q_i$ by $Y(\theta)$ if variable $x_i$ appears in positive form, or by $Y(-\theta)$ if the variable appears in negated form. Rotate the other qubits by the analogous rule.

\vspace{6pt}
3. Apply Q-NOR$_X$ (Table \ref{QnorX}) targeting an ancilla prepared in $\ket{1}$, and measure the ancilla in the $\ket{0}$, $\ket{1}$ basis. Abort if $\ket{0}$ is seen.

\vspace{6pt}
4. Reverse all three rotations applied in Step 2.
 }}\vspace{10pt}

If indeed $\theta=\frac{\pi}{2}$ then this is entirely equivalent to the na\"ive protocol, with the same low success probability i.e. we have probability $2^{-n}$ of `passing' all the clause checks, when we start from an initial state $\ket{++\dots +}$ and there is exactly one solution to the 3-SAT problem.

Generally when we {\it pass} a clause check operation, this leaves the register in a state with zero projection on the particular three-qubit state that {\it would have} failed (i.e. the state that would have rotated to $\ket{---}$). Asking for the state that passes all clause checks, for some particular value of $\theta$, is therefore a special case of the quantum $k$-SAT problem introduced by Bravyi~\cite{Bravyi06}.

It is helpful to think of a clause check process as checking the truth values in a basis defined by $\theta$. Suppose that a particular clause contains a boolean $a$ for which the corresponding qubit is $Q_a$. If the variable appears in direct form, then classically the clause would certainly evaluate to {\small TRUE} if boolean $a$ is itself {\small TRUE}. The equivalent statement for the quantum case is this: the clause check will certainly be {\it passed} if qubit $Q_a$ is in state $Y(-\theta)\ket{+}$. The reason is simple: a clause check process begins by applying rotation $Y(\theta)$ to $Q_a$, thus leaving it in state $\ket{+}$, and then it is impossible that the ancilla will flip during the Q-NOR$_X$ operation {\it regardless of the other two qubits associated with the clause}. Equivalently, if the clause contains the variable in its negated form $\lnot a$ then classically the clause will certainly be {\small TRUE} is $a$ is {\small FALSE}, while the quantum clause check will be passed if qubit $Q_a$ is in state $Y(\theta)\ket{+}$. The mapping is therefore,
\begin{eqnarray}
{\small \rm TRUE}\rightarrow Y(-\theta)\ket{+}&=&\cos\phi\ket{0}+\sin\phi\ket{1}\ \ \ \ \phi\equiv\frac{2\theta+\pi}{4}\nonumber\\
{\small \rm FALSE}\rightarrow \ \ Y(\theta)\ket{+}&=&\sin\phi\ket{0}+\cos\phi\ket{1}.
\label{nonOrthogBasis}
\end{eqnarray}
We see that in the limit $\theta=\frac{\pi}{2}$ we have the orthogonal states $\ket{1}$ and $\ket{0}$, whereas in the limit $\theta=0$ the truth values are degenerate: both correspond to $\ket{+}$ (so that passing all clause checks is inevitable for an initial state $\ket{++..+}$ but this is meaningless). 

For intermediate values of $\theta$ the states are non-orthogonal: We cannot use a measurement to determine which state a qubit is in with certainty. Moreover the various check operations need not {\it commute}: if two clause checks do not commute, then the order in which two clause checks are performed affects the resulting state of the qubit register. Two checks will not commute if they share at least one variable in common, and either (a) the two checks are using different $\theta$ values, or (b) they are using the same $\theta$ value but a variable appears in direct form in one clause and negated form in the other. That is, they do not commute if there is a conflict in the basis in which they are performing their projections. 

Consider the significance of non-commuting clause checks. If a clause check $\mathcal{C}$ is performed and is successfully passed, and then that same clause check $\mathcal{C}$ is immediately repeated, then the second instance must succeed (since the first check projected out, i.e. removed, any part of the overall state that could fail). This remains true if one or more additional clause checks which commute with $\mathcal{C}$ are performed in the interval between the two instance of $\mathcal{C}$. However if a non-commuting clause check $\mathcal{D}$ is performed between the two instances of $\mathcal{C}$, then the second instance of $\mathcal{C}$ is not guaranteed to succeed -- the act of passing though clause check $\mathcal{D}$ will (in general) have re-introduced a finite probability of failing $\mathcal{C}$. Now consider applying a {\it full cycle} of clause checks, i.e. one clause check for each clause of the 3-SAT, applied in some fixed order\footnote{In all cases I consider, I randomly select the order of the 3-SAT clauses at the start of an algorithm and then maintain that order through the algorithm}. The fact that the system's state has successfully passed one such complete cycle {\it does not imply} it would necessarily pass a second complete cycle - in fact it may fail the very first clause check of the second cycle, since several non-commuting checks will have been applied since that check was previously passed.

Suppose that we do indeed apply a {\it full cycle} of clause checks. Given that the 3-SAT problem has exactly one solution, then there is one special state of the qubit register which is guaranteed to pass all checks. This is simply the following state, where I use the bold symbol in $\ket{\bm \theta}$ to denote the entire $n-$qubit register:
\begin{equation}
\ket{\bm \theta}=\bigotimes_i \ Y\left(L_i\theta\right)\ket{+}_i 
\label{fullTheta}
\end{equation}
where $L_i=-1$ if the satisfying logical value of the boolean variable $x_i$ is {\small TRUE}, and $L_i=+1$ if that value is $x_i$ is {\small FALSE}. Any clause check will then rotate at least one of its three qubits to the $\ket{+}$ state, and therefore that check will succeed (since only the $\ket{---}$ row in Table~\ref{QnorX} leads to a clause check failing). A converse argument shows that {\em only} this state can satisfy all clause checks with certainty\footnote{To see the uniqueness of this state, consider the $2^n$ states of the form of Eqn.~(\ref{fullTheta}), one for each of the possible sets of boolean assignments, i.e. each possible set of $L_i$ values. These states are independent, and therefore although they are not orthogonal for general $\theta$ nevertheless they suffice as a basis to express any state of the qubit register. Decompose any general state in this way and then note that each element will fail at least one clause check, except for the element corresponding to the boolean assignment that satisfies our single-solution 3-SAT problem.}, and this observation is verified for every $N_S=1$ problem for which I've performed a numerical simulation (hundreds of thousands of cases). 

This raises some interesting possibilities. If the state given in Eqn.~\ref{fullTheta} can be created, for any $\theta>0$, then the qubit register encodes the satisfying solution to the 3-SAT problem. However in general one cannot deterministically obtain this information by directly measuring the register qubit-by-qubit. This is possible only in the limit $\theta=\frac{\pi}{2}$, where one can simply measure in the $\ket{0}$, $\ket{1}$ basis to determine each boolean value {\small FALSE} or {\small TRUE}; but of course, simply setting $\theta=\frac{\pi}{2}$, initialising to $\ket{++..+}$, and performing clause checks will lead to the low probability $2^{-n}$ of passing all clause checks. Alternatively we could start by performing clause checks with some $\theta$ close to zero (and which will therefore succeed with high probability) and then slowly increase $\theta$, so as to gradually transform the qubit register toward the desired final state with $\theta=\frac{\pi}{2}$. We might hope that the probability of successfully reaching that state is much greater than $2^{-n}$; this would be analogous to adiabatic quantum computing, where the parameters of the physical system are slowly changed and one hopes that the system will remain in its ground state even as the nature of that state changes. 

\bigskip

\noindent {\bf\large An ``adiabatic-like" approach}

\smallskip

\begin{figure}[t]
\centering
\includegraphics[width=.99\linewidth]{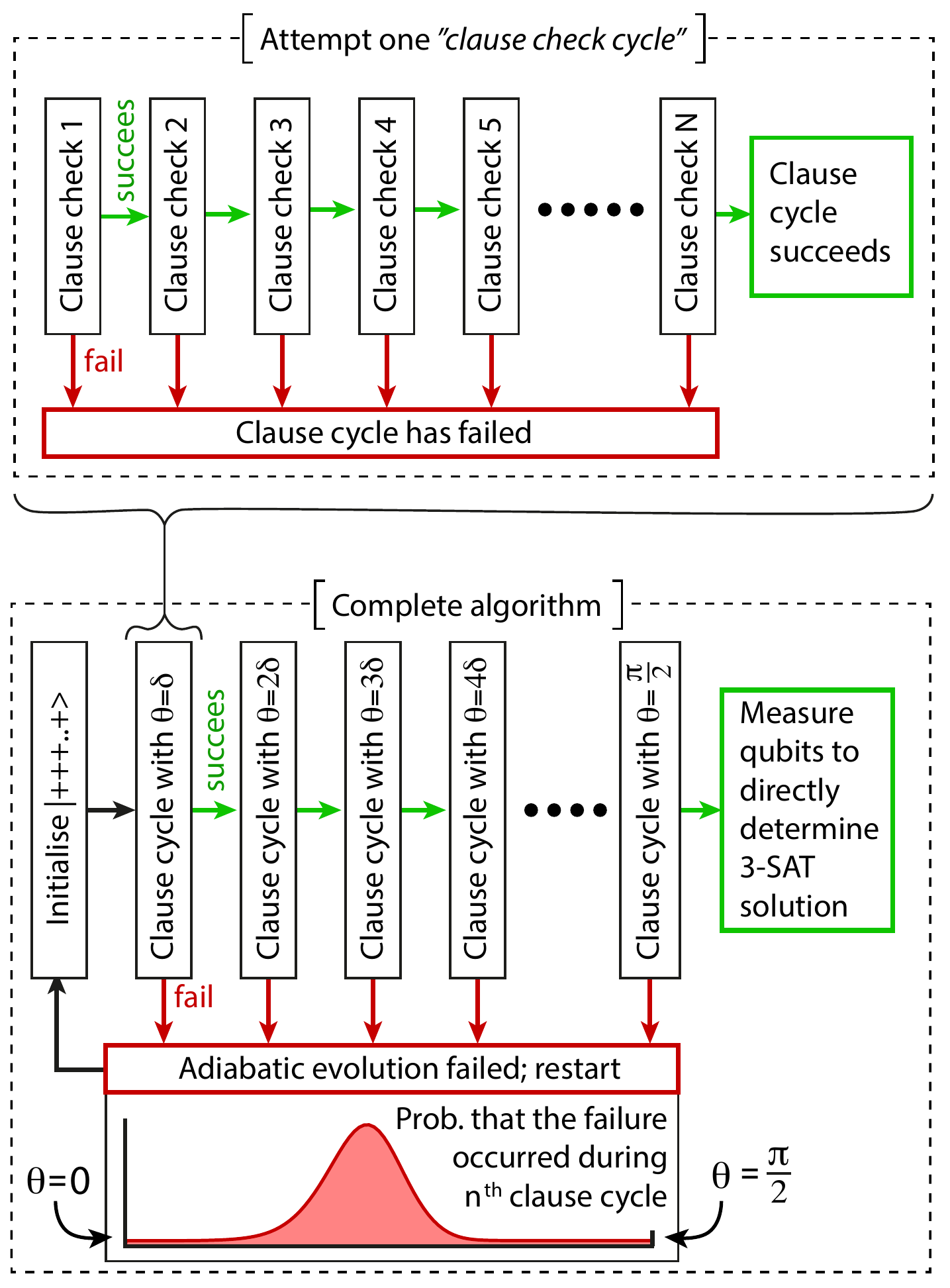}
\caption{ Flow diagram for the basic `adiabatic-like' algorithm. 
\label{fig:alg_flow}
}
\end{figure}

A small change in the parameter $\theta$ corresponds to a small change in the state $\ket{\bm\theta}$ given by Eqn.~\ref{fullTheta}, in fact $|\langle{\bm\theta^\prime}\ket{\bm \theta}|^2\approx 1-\frac{n}{4}\epsilon^2$ if $\epsilon\equiv\theta^\prime-\theta$ is small (c.f. the Zeno effect), and we may therefore hope to be able to guide the system to the ideal state $\ket{{\bm\theta}=\frac{\pi}{2}}$ and arrive there with a probability that approaches unity for sufficiently slow evolution of $\theta$. This intuition will prove to be correct, but with the significant caveat that the probability of failure can only be made small for {\it extremely} slow evolution. We can select any rate of change of $\theta$ simply by introducing an appropriate number of clause checks -- in my simulations of this approach I set parameter $\theta$ to a given value, perform one complete cycle of clause checks, and then increment $\theta$ and repeat. See Fig.~\ref{fig:alg_flow}.

\begin{figure}[!b]
\centering
\includegraphics[width=.8\linewidth]{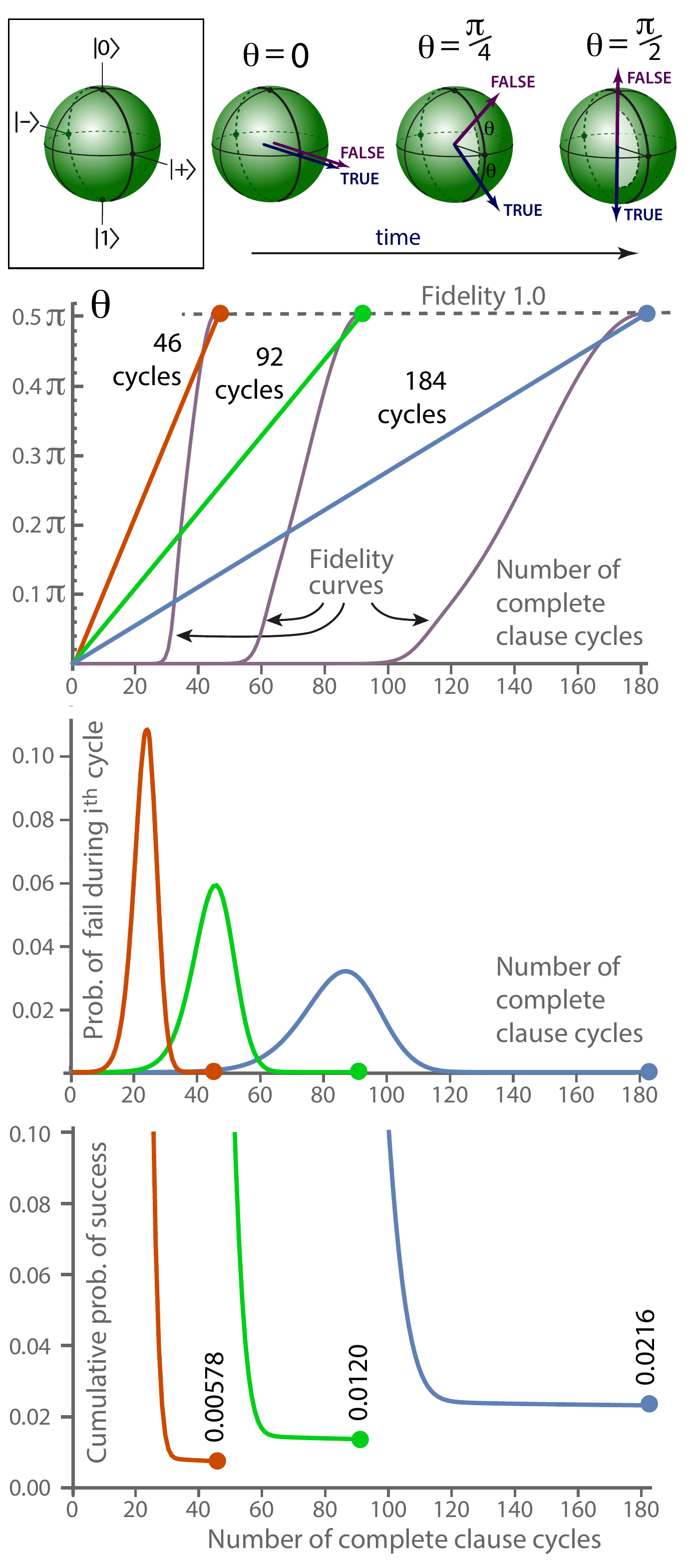}
\caption{Data from a simulation of the `adiabatic-like' approach (Fig.~\ref{fig:alg_flow}) with a 3-SAT problem of $24$ booleans and a unique solution. The red, green and blue traces correspond to three different rates of {\it linear} evolution of the parameter $\theta$. The slower the evolution, the higher the chance of reaching the desired state ultimate state $\ket{{\bm\theta}=\frac{\pi}{2}}$; the fidelity with respect to this target is shown by the grey curves in the uppermost figure. However it is {\it not} optimal to go very slowly, as shown by the cost analysis in Fig.~\ref{fig:simpleLinearPart2}. }
\label{fig:simpleLinearPart1}
\end{figure}

Suppose that we try increasing $\theta$ as a linear function of time, i.e. incrementing $\theta$ by a small {\em fixed} amount $\delta$ after each cycle. What is the optimum value of $\delta$? One might imagine that it is best to make $\delta$ as small as is necessary to make the probability of success approach unity. This intuition proves to be incorrect. To understand why, it is instructive to look at a simulation of a particular problem; in Fig.~\ref{fig:simpleLinearPart1} I show data for a typical USA 3-SAT problem involving $24$ booleans. We see that, as expected, a slower evolution leads to a better chance of reaching the desired $\theta=\frac{\pi}{2}$ final state.

\begin{figure}[!b]
\centering
\includegraphics[width=.9\linewidth]{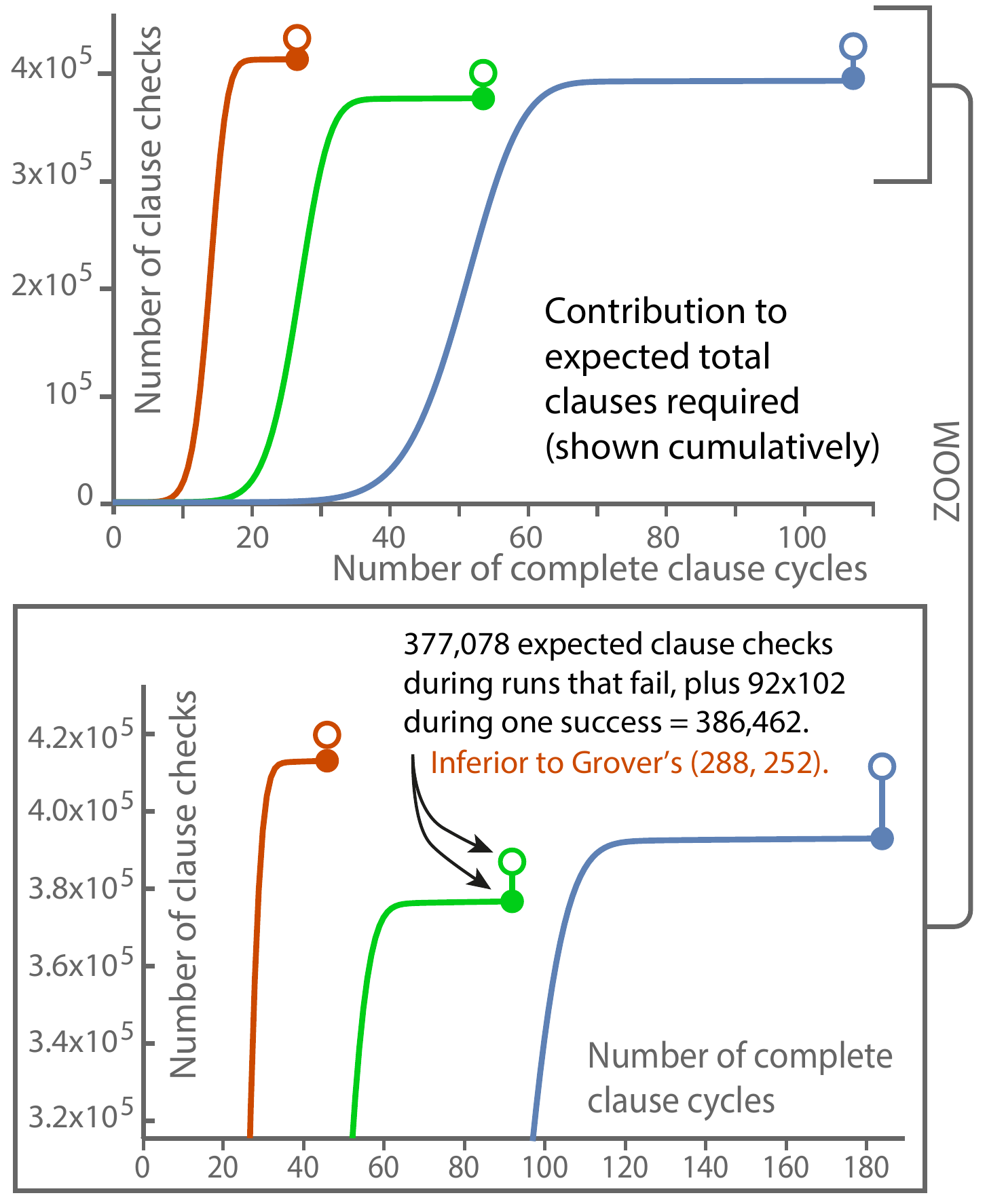}
\caption{A cost analysis for the same problem considered in Fig.~\ref{fig:simpleLinearPart1}. Curves show the expected {\em total} number of clause checks that are performed during runs that ultimately fail, assuming that we try enough times to achieve one success (and therefore reveal the solution to the 3-SAT). The curves are cumulative, showing the contribution to this total made at different stages (so, even though the cost of a failure late in the evolution is high, the contribution to the expected total is small because such failures are rare). By adding this expected total to the number of clause checks in one successful run ($96\times102=9,792$ for the green trace), we can find the expected number of clause checks required to solve this 3-SAT. This total of $386,462$ is optimal for a linear ramping of $\theta$, but other ramps are better; see Table~\ref{dataTable} and Fig.~\ref{compare3}. }
\label{fig:simpleLinearPart2}
\end{figure}

However, in fact one should not evolve more slowly than a certain optimum, and this optimum corresponds to a success probability that is still far below unity. This is because there is a peak to the failure probability when $\theta$ is already substantial, around in the mid-point in the evolution as can be seen from the middle graph of Fig.~\ref{fig:simpleLinearPart1}. The lowest graph in Fig.~\ref{fig:simpleLinearPart1} shows that doubling the number of increments from $92$ (green) to $184$ (blue) causes the success probability for a given `run' to increase. However the expected number of steps-until-failure on a each failed run {\it increases by a similar factor} because the peak probability of failure is now occurring after about twice as many cycles. The net consequence is revealed in Fig.~\ref{fig:simpleLinearPart2} where we see that going from green to blue corresponds to a slight {\em increase} in overall expected number of clause checks required before a 3-SAT solution is found.

\begin{figure}[b]
\centering
\includegraphics[width=.9\linewidth]{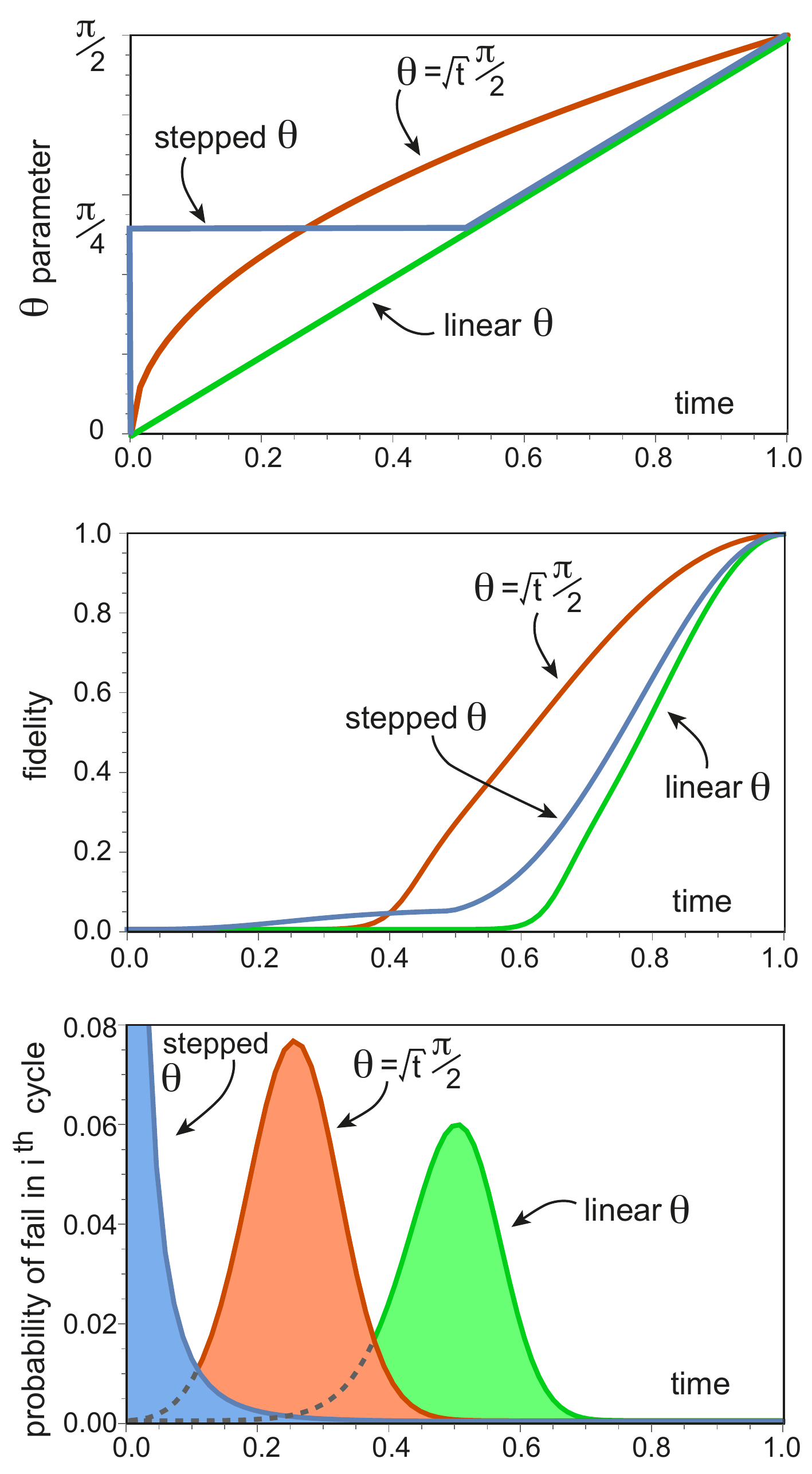}
\caption{A comparison of the three variations on the measurement-driven algorithm used to solve a particular $24$ boolean USA 3-SAT problem which was analysed earlier in Figs.~\ref{fig:simpleLinearPart1} \& \ref{fig:simpleLinearPart2}. The green curve here corresponds to the green curves in those earlier figures. The parameters and performance of these three approaches are summarised in Table~\ref{compareDataTable}. The best performance is obtained using the hybrid approach, i.e. the `stepped $\theta$' which starts from $\theta=(0.56)\frac{\pi}{2}$ and holds this fixed for $37$ full clause cycles before linearly ramping to $\theta=\frac{\pi}{2}$ over the remaining cycles.\\
}
\label{compare3}
\end{figure}

It is interesting to compare the performance of this simple measurement-driven algorithm to a direct use of Grover's search algorithm. For the particular 3-SAT problem used for the simulations depicted in Figs.~\ref{fig:simpleLinearPart1} and \ref{fig:simpleLinearPart2}, the overall performance with the optimum choice of $92$ full clause check cycles, is somewhat inferior to Grover's algorithm. The expected total number of clause tests in Grover's would be $288,252$ (see Appendix I), as compared to $386,462$ for the present algorithm. If however we alter the ramping function we can quickly improve the performance of the present approach; by using  $\theta=\frac{\pi}{2}\sqrt{c/c_{\rm tot}}$, for the $c^{\rm th}$ cycle, we reach an expected number of clauses of only $190,885$ (this root-$t$ evolution is shown later in Fig.~\ref{compare3}). There are a number of considerations that mean this comparison to Grover's algorithm has limited meaning -- for example, the performance of Grover's is the same for all 3-SAT problems of a given size and with a given number of solutions, whereas for the present approach there is variation (as discussed presently) and the data depicted in Figs.~\ref{fig:simpleLinearPart1} \& \ref{fig:simpleLinearPart2} is merely typical rather than definitive. Nevertheless is it interesting that the performance of Grover's algorithm can be matched, roughly speaking, since the present measurement-driven approach is apparently very different. Moreover by adjusting the present approach its performance can be considerably improved.

Although there is an analogy between the slow evolution of $\theta$ used here and the slow change of a Hamiltonian in AQC, already we are exploiting a feature which AQC does not offer: the moment that a run has failed, this is known and we can restart immediately without `throwing good money after bad'. Indeed, this is the reason that the square-root ramping algorithm outperforms the linear one -- it `front loads' the failure probability more aggressively, thus allowing us to fail faster! 

There are a number of approaches one might consider to refine this adiabatic-like approach, including more complex evolution of $\theta$ over the course of the process  or even attempting to `repair' a state after a clause check fails. However I will not consider these further here, but instead I focus on simply reducing the expected number of steps in a failed run by `front loading' the failure probability even more aggressively. This approach will move away from the idea of an `adiabatic' evolution, since we will now see a substantial failure probability for each clause check early in the process.  

\bigskip

\noindent{\bf \large A ``sculpting" approach}

\smallskip

\begin{figure}[b!]
\centering
\includegraphics[width=.95\linewidth]{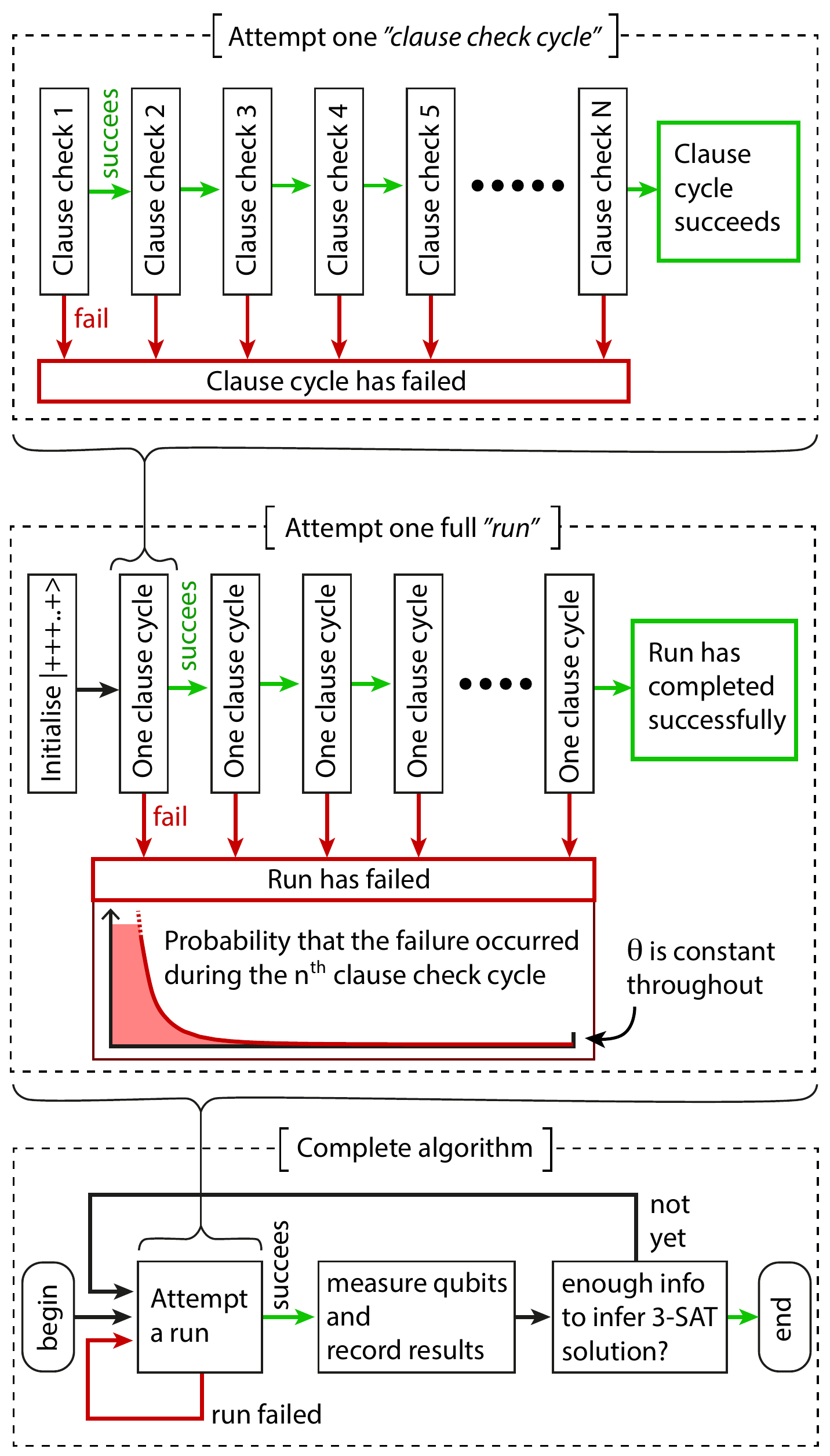}
\caption{ The `sculpting' algorithm aggressively `front loads' the failure probability.  }
\label{fig:alg_sculpt}
\end{figure}

Consider an approach where we initialise all qubits in $\ket{+}$, select some fixed value of $\theta$ (well above zero, yet well short of the ideal $\theta=\frac{\pi}{2}$ state) and simply begin performing clause checks. The clause checks early in the process will be the most likely to fail, and (if passed) will have the greatest effect on the system's state -- somewhat as a sculptor might begin by carving away large pieces of stone. Our aim is to create state $\ket{\bm \theta}$ to a high fidelity after a sufficient number of ``passed" clause checks. Because of the non-orthogonality of the clause check operations, it  may require many complete clause check cycles to reach high fidelity. As I presently discuss, although this target state is not the ideal $\theta=\frac{\pi}{2}$ state, we can extract useful information from it and then repeat, as shown in Fig.~\ref{fig:alg_sculpt}. Alternatively, having achieved the state $\ket{\bm\theta}$ we could {\it then} apply an adiabatic-like evolution as a kind of secondary phase, in which we guide the system onward from this $\theta$ to the ideal $\frac{\pi}{2}$, as shown in Fig.~\ref{fig:2phase}.

\bigskip

Since there is a finite projection between the initial state and the target state $\ket{\bm \theta}$, we expect to to arrive at this target with at least probability 
\begin{equation}
p=|\langle {\bm \theta} | ++\dots +\rangle|^2=\left(\cos\frac{\theta}{2}\right)^{2n}.
\label{probRunSuccess}
\end{equation}
However since we are selecting a $\theta$ value that is substantially greater than zero, the process is not an adiabatic evolution and we should not expect to achieve a success probability that is significantly higher than this bound.
The simulations I have performed verify this. 

It is interesting to monitor the fidelity of the qubit register with respect to the target state $\ket{\bm\theta}$ as the algorithm proceeds, i.e. as each successive clause check operation is successfully passed. Specifically, it is useful to note how many clause checks must be passed before the fidelity reaches a given threshold (since the fidelity only asymptotically approaches unity, we must select some threshold below unity as a target). Having chosen a target fidelity, one can ask whether different 3-SAT problems take longer to reach the target and moreover how this varies as the number of booleans increases. 

\begin{figure}[b]
\centering
\includegraphics[width=.7\linewidth]{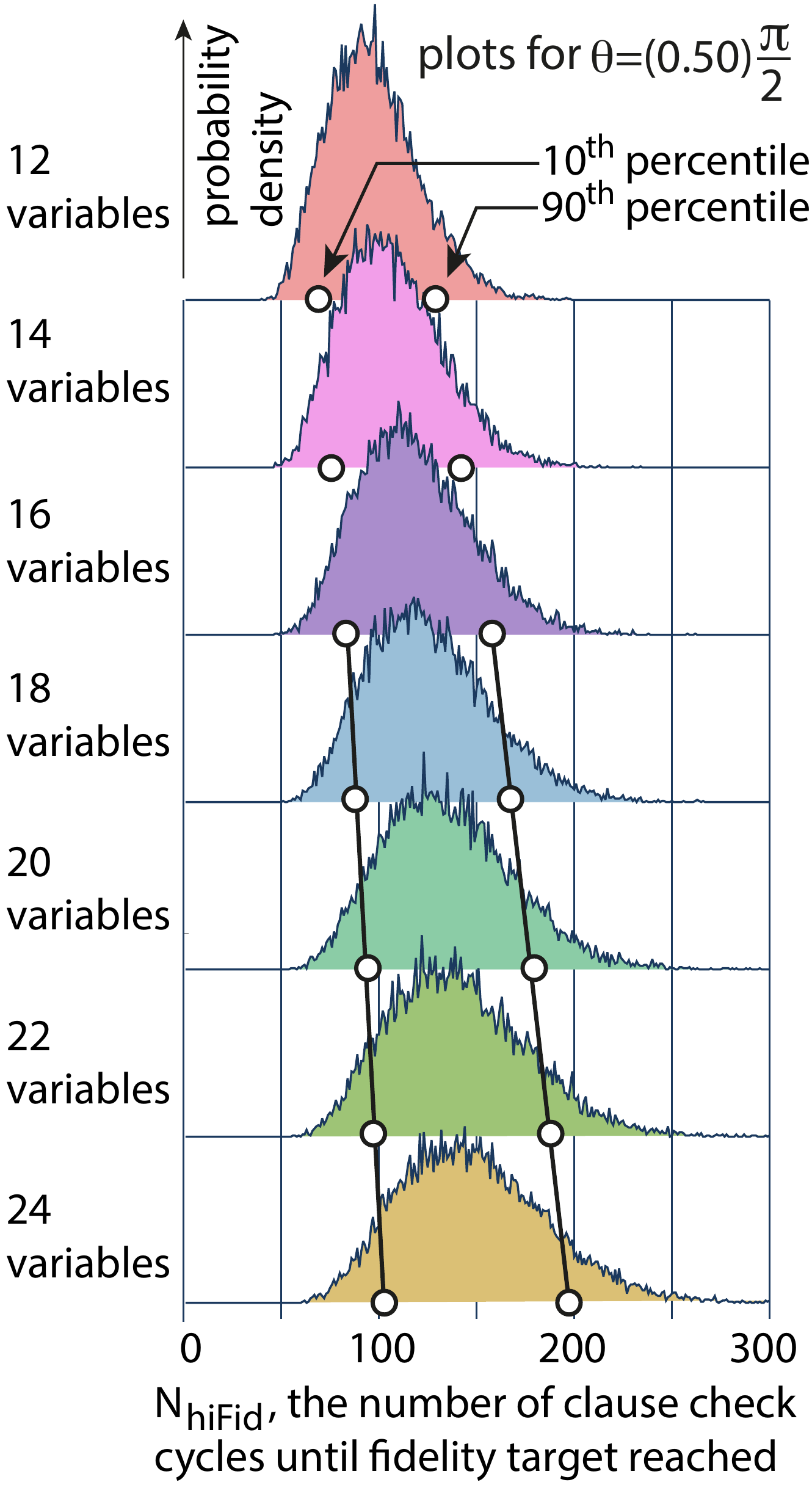}
\caption{ For a given number of booleans, a randomly selected 3-SAT instance (with a unique solution) may correspond to a relatively quickly evolving quantum system or a rather slower one -- i.e. there is {\em variation} in the rate at which the quantum state evolves to a target state $\ket{\bm \theta}$ according to the specific nature of the problem . I have not studied the dependence in detail but, for example, having multiple solutions {\it near} the unique correct solution tends to slow it down. As the problem size (number of booleans) increases then this distribution changes -- but over the range I can inspect the distribution spreads only modestly. \\
The right-side tails of these distributions appear to be `heavier' than $\exp(-kN^2)$, and are reasonably fitted by $\exp(-kN)$. }
\label{fig:timeTo999}
\end{figure}

Figure~\ref{fig:timeTo999} shows $N_{\rm hiFid}$, the total number of successful clause checks that must be performed when we start from the initial state $\ket{++..+}$ and aim to reach $\ket{\bm\theta}$ with a fidelity on the target state of $0.999$. This is, $N_{\rm hiFid}$ is number of clause checks until
\[
|\langle{\bm\theta}\ket{\Psi}|^2\geq0.999
\]
where $\ket{\Psi}$ is the instantaneous state of the qubit register at some point during the algorithm. The interesting observation is that there is a spread in $N_{\rm hiFid}$ even for a fixed choice of $n$ and $\theta$. Note that all problems generated for these simulations had a unique satisfying solution.

Obviously, if we were using this algorithm to seek the as-yet unknown solution to a 3-SAT problem, we would not know in advance whether this problem would prove to be at the `fast' or `slow' end of the spectrum. (Indeed, we would not know whether there {\it is} a solution and whether it is unique; I consider this point later, but suppose for the moment that it is known that any solution is indeed unique.) We would then need to make a choice of how many clause checks to require before calling a run a success and measuring the qubits, as called for by the overarching algorithm Fig.~\ref{fig:alg_sculpt}. I will use the symbol $N_{\rm full}$ for this user-selected number. A simple strategy would be just to choose $N_{\rm full}$ to large enough that the bulk of the distribution of $N_{\rm hiRes}$, say $99.9\%$ of it, is lower. Fortunately there is relatively little cost in setting a high value for $N_{\rm full}$, because once a high fidelity approximation to $\ket{\bm\theta}$ is reached then of course additional clause checks are unlikely to fail (but see the effect of imperfectly applied operations, discussed later). As for the possibility that the 3-SAT problem in question is an exceptionally `slow' one, beyond even our supposed $99.9\%$ threshold, assessing the impact of this lies beyond the range of the simulations and analysis I present here -- it may be that the distribution of $N_{\rm hiRes}$ has a hard upper limit so that this problem does not arise; conversely it may be the case that there is a signature in the statistics of measured qubits which can indicate a solution exists but that  $N_{\rm full}$ should be increased in order to find it.


The `sculpting' algorithm shown in Fig.~\ref{fig:alg_sculpt} calls for us to repeatedly create an approximation to $\ket{\bm\theta}$ and measure it. Each successful create-and-measure process will yield information that is correlated to the 3-SAT solution (as discussed later). But what is the expected number of clause checks $C_{\rm total}$ needed for one successful full run? We can write
\[
C_{\rm total}=R_{\rm RPS}\ C_{\rm CPR}
\]
where $R_{\rm RPS}$ is the expected number of runs that must be attempted in order to achieve one successful run (``RPS" for Runs Per Successful run), and $C_{\rm CPR}$ is the expected number of clause check operations per run (``CPR'' for Clauses Per Run). Now the former is simply $R_{\rm RPS}=(p_{\rm succeed})^{-1}$ where  $p_{\rm succeed}$ is the probability of a run succeeding, and as noted earlier this is well approximated by $p_{\rm succeed}\sim(\cos\frac{\theta}{2})^{2n}$ for $n$ qubits. The latter term can be written as
\[
C_{\rm CPR}=N_{\rm full}\ p_{\rm succeed}+\sum_{i=1}^{N_{\rm full}} i \ p_{\rm fail}(i)
\]
where $N_{\rm full}$ was introduced above as the user-selected number of clause checks in a successful complete run, and $p_{\rm fail}(i)$ is the probability that a run aborts specifically during the $i^{\rm th}$ clause check. Therefore

\begin{equation}
C_{\rm total}=N_{\rm full} +F\left(\sec^2\frac{\theta}{2}\right)^n\ \ \ {\rm where}\ \ \ F=\sum_{i=1}^{N_{\rm full}} i \ p_{\rm fail}(i).
\label{fullCostEqn}
\end{equation}

\begin{figure}[b]
\centering
\includegraphics[width=1.\linewidth]{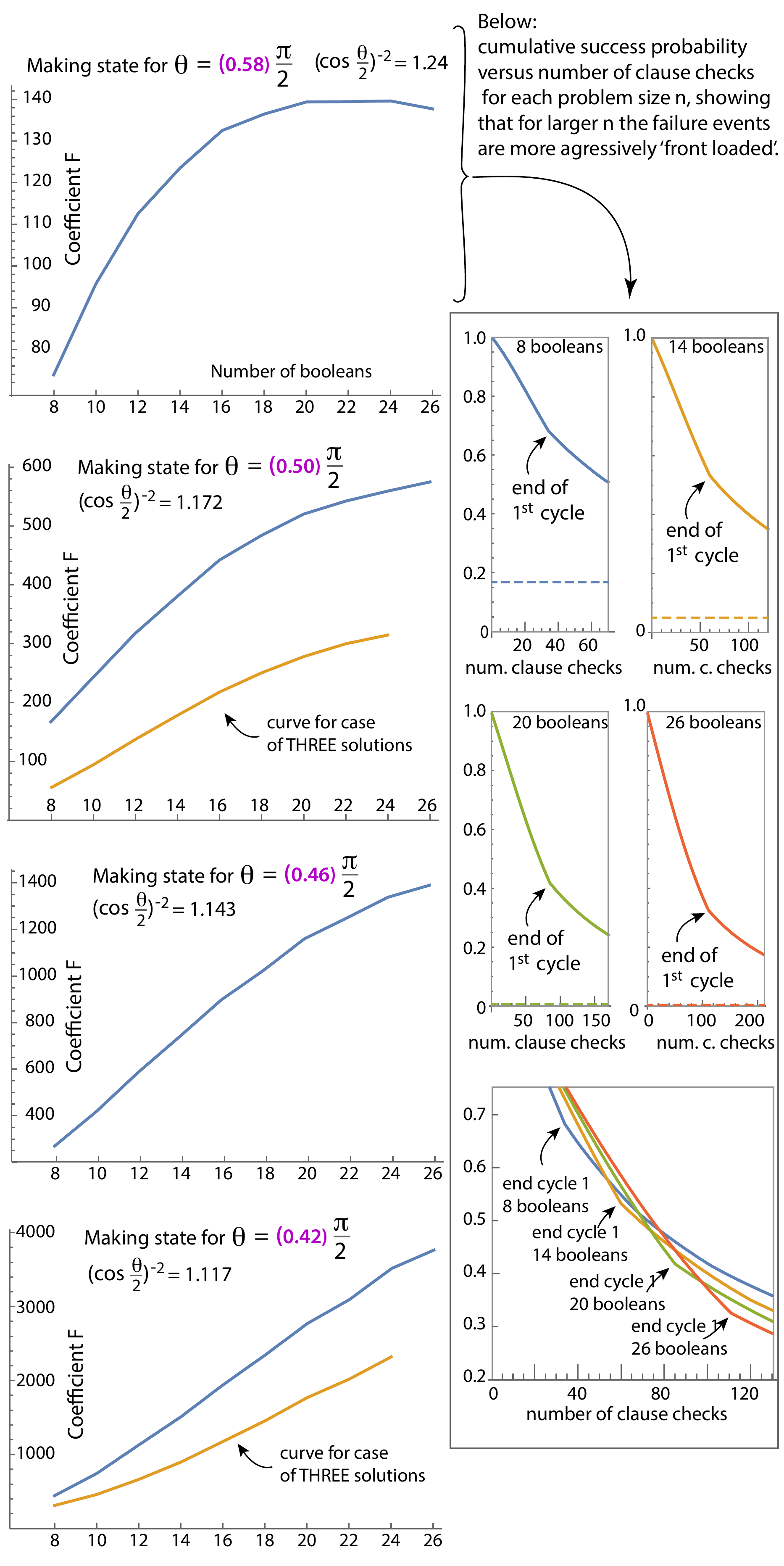}
\caption{Values of $F$ in Eqn.~\ref{fullCostEqn} determined from numerical experiments. Two curves for three solution cases are shown.
 } 
\label{fig:prefactors}
\end{figure}

As noted in Fig.~\ref{fig:timeTo999} and its caption, over the limited range for which I have performed numerics, the suitable choice for $N_{\rm full}$ appears to scale only linearly with $n$, the number of booleans.\footnote{In fact $N_{\rm full}$ must scale at least linearly, since we cannot create the solution-encoding state $\ket{\bm\theta}$ without having evaluated every clause at least once, and for non-trivial 3-SATs the number of clauses scales linearly with $n$.} Taking this to be the case, then the scaling for $F$ must be at-most linear, since it is the expected number of clause checks in a run, given that {\it the run fails} i.e. it does not achieve $N_{\rm full}$ clause checks. Figure~\ref{fig:prefactors} shows how $F$ behaves over a range of problem sizes. One sees a linear dependence becoming sub-linear for large $n$, and even declining as $n$ becomes very large.

If the trends appearing in the numerics of Fig.~\ref{fig:timeTo999} and Fig.~\ref{fig:prefactors} prove to continue to large $n$, in particular if the required $N_{\rm full}$ continues to scale only linearly with $n$, then this algorithm constitutes a means of creating a high fidelity approximation to state $\ket{\bm\theta}$ in a time scaling as $\left(\sec^2(\frac{\theta}{2})\right)^n$ where we can choose $\theta$ in the range $0<\theta<\frac{\pi}{2}$. But what use is a single instance of $\ket{\bm\theta}$ in solving the 3-SAT? Recall that if $\theta=\frac{\pi}{2}$ then the state $\ket{\bm\theta}$ can simply be measured to reveal the (unique) solution; but of course then $\sec^2(\frac{\theta}{2})=2$ and we have merely matched the scaling of the time cost for basic random classical search. In general if we measure the qubits in a perfectly-prepared state $\ket{\bm\theta}$ in the $z-$basis then each qubit (independently of the others) gives the `correct' truth value with probability $\frac{1}{2}(1+\sin\theta)$. Thus by repeatedly generating, and measuring out, the state $\ket{\bm\theta}$ we can rapidly reach a high degree of certainty as to the 3-SAT solution. An analysis is presented in Appendix 4; in short, the expected number of repetitions $R$ needed to infer the full 3-SAT solution grows only as a logarithmically with $n$. 

Now in practice we cannot create a perfect $\ket{\bm\theta}$ in finite time for $\theta<\frac{\pi}{2}$, however as shown in Fig.~\ref{fig:timeTo999} we can reach a high fidelity for the entire state, and then each qubit's probability of correlating with the correct truth value will indeed approach the ideal of $\frac{1}{2}(1+\sin\theta)$.

\begin{figure}[b]
\centering
\includegraphics[width=.95\linewidth]{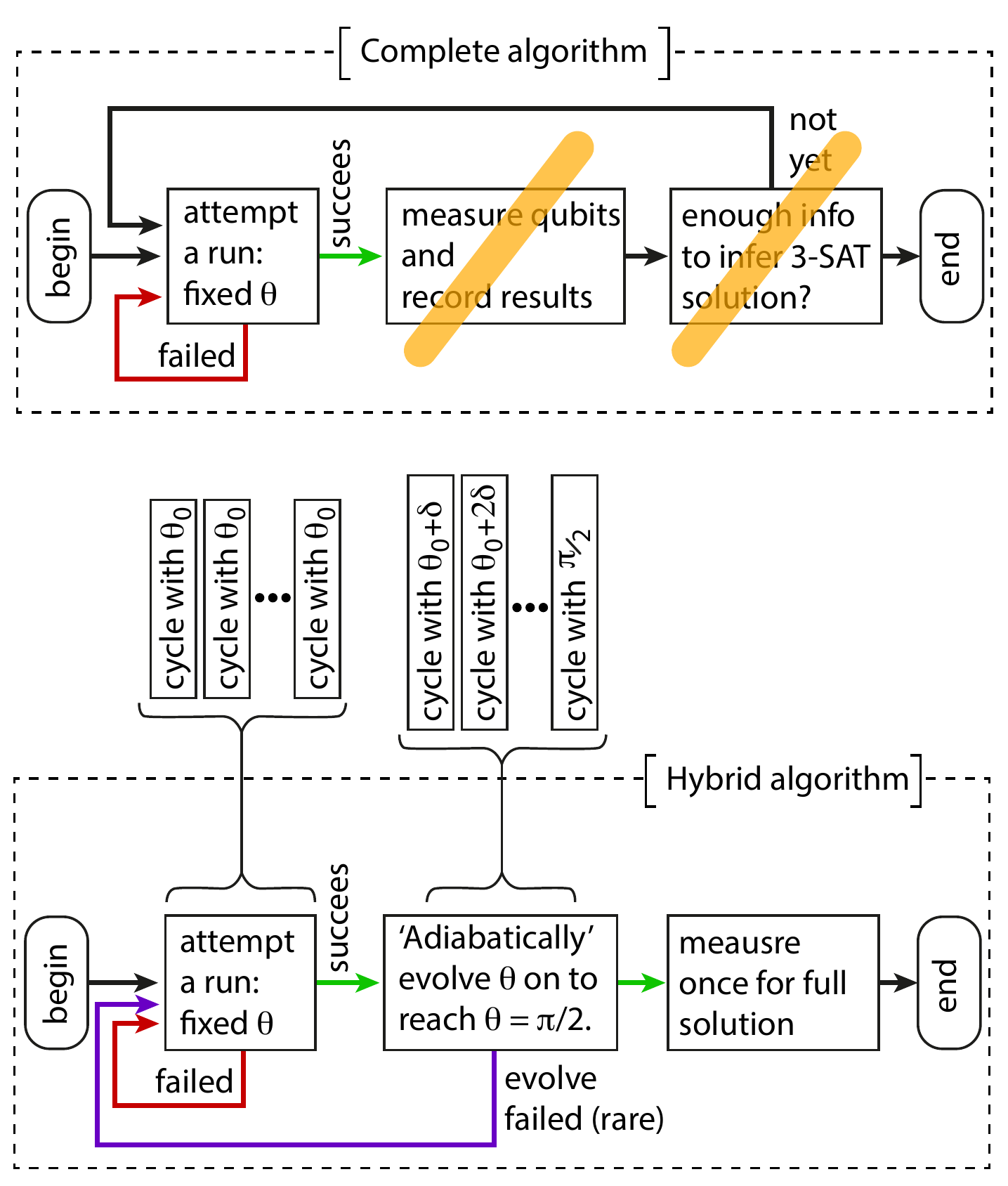}
\caption{A hybrid algorithm. Instead of the repeating prepare-and-measure process in the sculpting algorithm (upper panel here)  we need only `sculpt' $\ket{\bm\theta}$ once, using some intermediate value of $\theta=\theta_0$, and we proceed to evolve slowly to $\theta=\frac{\pi}{2}$. }
\label{fig:hybrid}
\end{figure}

\bigskip

\noindent{\bf \large A hybrid approach}

\smallskip

The previous two sections have described two approaches to measurement-driven 3-SAT solving: an `adiabatic-like' approach where the key parameter $\theta$ is slowly evolved, and a `sculpting' approach where $\theta=\theta_0$ is fixed and consequently the system's state changes dramatically at first and then is more slowly honed. A drawback of the latter is that we need to repeatedly create-and-measure $\ket{\bm\psi}$ in order to glean enough information to solve the 3-SAT. 

\begin{figure}[b]
\centering
\includegraphics[width=.9\linewidth]{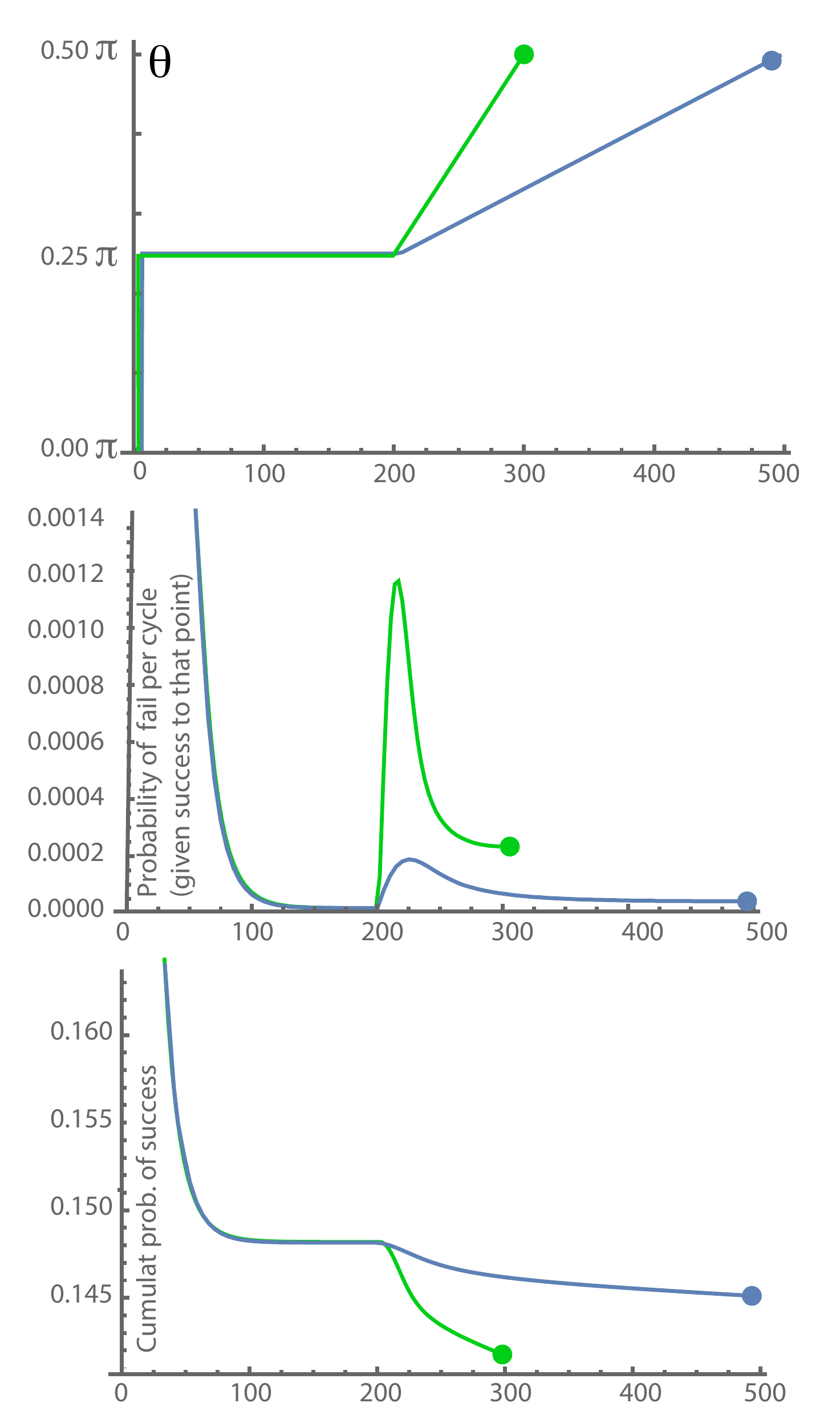}
\caption{Hybrid algorithm with $\theta_{0}=\frac{\pi}{2}$. By moving sufficiently slowly, the failure probability in the second phase is made very small.}
\label{fig:2phaseData}
\end{figure}

A third possibility is to create the state $\ket{\bm\theta}$ for some $\theta_0<\frac{\pi}{2}$ via the sculpting approach, and then continue with a slow evolution to $\theta=\frac{\pi}{2}$ by the adiabatic-like approach (see Fig.~\ref{fig:hybrid}). Then we gain the entire 3-SAT solution by measuring the register once, at the end of the first successful run of the algorithm. 

Whether or not this approach is workable depends on the cost of the second phase: inevitably there will be some probability of failure during this phase, and (as we saw from the analysis of the original adiabatic-like approach) the {\it cost} of this failure is not necessarily reduced when the evolution is slowed. However, at least for some values of $\theta_0$ the hybrid approach does indeed succeed, as shown in Fig.~\ref{fig:2phaseData} and Fig.~\ref{compare3}.

In Figure~\ref{compare3} three measurement-driven approaches compared. Table~\ref{compareDataTable}  summaries the performance, with the caveat that there are further optimisations possible for the `hybrid' approach (I made the arbitrary choice to split the evolution `50/50' between the static $\theta$ phase and the ramping phase, and to use a linear ramp). 

\begin{table}[!h]
\caption{Comparison of approaches}
\centering
\begin{tabular}{|l|r|r|r|}
\hline
Approach &  Optimal & Tries  & Total \\
Used &  Cycles & Needed  & Time \\
\hline
Grover's Algorithm & N/A & 1.18\footnote{Tries is $1.18$ rather than $1$, see Appendix on Grover's} & 288,252\\
Adiabatic-like: Linear ramp & 92 & 83 & 386,462 \\
Adiabatic-like: Sqr-root ramp\ \  & 73 & 99 & 190,885 \\
Hybrid with `stepped' $\theta$ ramp\footnote{Using $\theta_0=(0.56)\pi/2$ which is optimal for this problem.}& 75=37+38 & 143 &38,031\\
\hline
\end{tabular}
\label{compareDataTable}
\end{table}

For each approach, the table lists under `Cycles' the number of complete clause check cycles that were found to be optimal in solving a particular $24$ boolean 3-SAT problem (the same problem used in Fig.~\ref{fig:simpleLinearPart1}). This is the user-selected number that determines how slowly $\theta$ evolves to $\frac{\pi}{2}$. However most attempts will fail; consequently we expect to have to try multiple times and this is listed under `Tries'. But the important figure is the one in the rightmost column: how many clause checks do we expect to have to perform {\em in total} before we eventually have a successful run and therefore find the 3-SAT solution.

\section{Multiple 3-SAT solutions}

To this point, I have considered only 3-SAT problems that have a unique satisfying assignment (USA). The extension to cases with multiple solutions is straightforward. For the USA case, for a given $\theta$ there was a single state that passed all the stabiliser checks with certainty, i.e. Eqn. (\ref{fullTheta}). Now instead there is a subspace spanned by a set of such states, one for each solution of the 3-SAT. These states are not orthogonal (except when $\theta=\frac{\pi}{2}$), but they are mutually independent. 

The goal of the algorithm will now be to drive the qubit register into this subspace. There will be no guarantee as to whether the final state contains equal amplitudes of the different solutions, and indeed in general it will not; however provided that we can learn one of the solutions then we can modify the 3-SAT problem to exclude this solution (for example by adding a clause which the found solution does not satisfy) and thus search for others.\footnote{A difficult case would be a 3-SAT with two solutions differing in only one boolean, so that adding a clause to exclude the found solution would likely exclude the alternative solution also. In that case Order$(n/3)$ further experiments would suffice to find the second solution, by trying a series of different additional clauses.
}

One can obtain a fidelity metric to measure how well our qubit register is driven towards the satisfying subspace (see Appendix 2), and thus we can ask how long it takes to reach the same $\sim0.999$ pre-qubit fidelity criterion that we used earlier for the USA problems. Of course, now only the qubits that correspond to booleans whose value is the same over all solutions will approach a specific state, and as we might expect the algorithm runs a little faster. A couple of such plots are presented in \ref{fig:prefactors}, for comparison with unique solution cases. 

Obtaining one full solution is straightforward if we are taking the simple adiabatic-like approach (Fig.~\ref{fig:alg_flow}) or the hybrid `two phase' approach (Fig~\ref{fig:2phaseData}) which both conclude with $\theta=\frac{\pi}{2}$, since then merely measuring all qubits in the $z$-basis will reveal a solution. 

If instead we are taking the sculpting route (Fig.~\ref{fig:alg_sculpt}) of preparing $\ket{\bm\theta}$ for some $\theta>0$ and then measuring for partial information, before repeating for additional information, then things are a little more complex. One approach would be as follows: In a multi-solution 3-SAT problem whenever a given boolean varies between the solutions (i.e. it is {\small TRUE} for some solution(s) but not all solutions) then the corresponding qubits will show a lesser asymmetry (frequency of $0$'s versus $1$'s) as compared to the expected bias $~(1+\sin\theta)/2$. Noticing this, one would reduce the SAT problem by setting the corresponding boolean to {\small TRUE}, say, in order to narrow the search for a satisfying solution. This would be repeated until the qubits in the reduced problem show the proper correlation to unique truth vales. 

\begin{figure}[b]
\centering
\includegraphics[width=.85\linewidth]{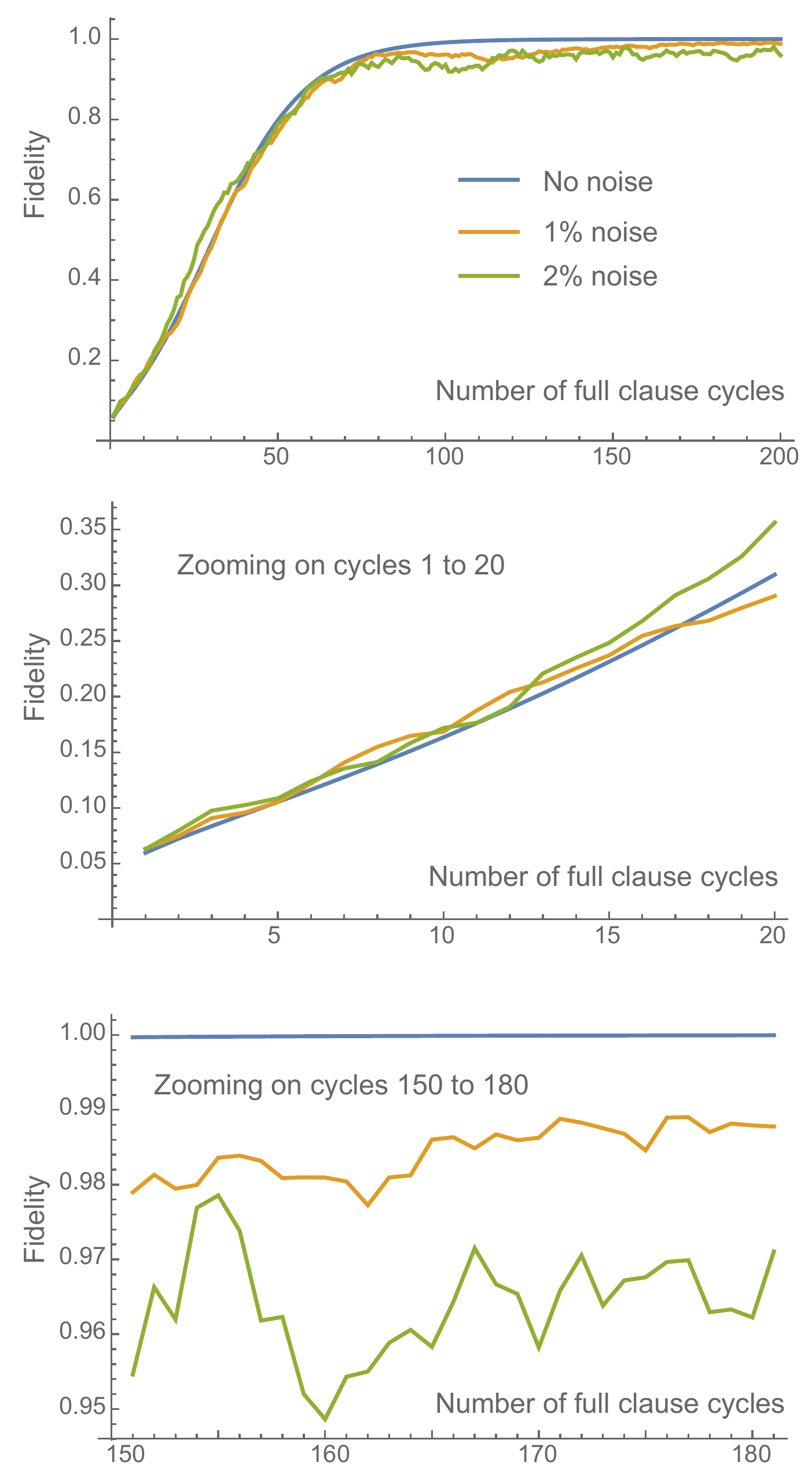}
\caption{The effect of adding noise to the rotations during the running of the `sculpting' algorithm ($\theta=(0.5)\frac{\pi}{2}$) on a particular 20 boolean problem. Note that noise can boost the fidelity (the green line goes above the blue in places).\\
}
\label{noiseGraphs}
\end{figure}

\section{Robustness versus errors}

There is reason to hope that the approach described here can have a degree of {\it inherent} error tolerance, where by `error' I mean imperfections in the preparation, rotation, and measurement of qubits, or general decoherence to the qubits from the environment. Because the approach works by continually projecting the system toward a target state $\ket{\bm \theta}$, certain small deviations to the state may be corrected. Meanwhile large deviations may cause the protocol to abort, but that is also a form of error handling.

As a preliminary exploration of this possibility, I introduced a small random contribution to every single-qubit rotation in the sculpting algorithm, i.e. each rotation over or under rotates by a random defect that is capped at a certain percent. The results are shown in Fig.~\ref{noiseGraphs}. One can see that indeed the algorithm still functions, albeit at a slightly reduced level.

A proper exploration of error tolerance would of course require far more that this -- general single-qubit errors (rather than merely over/under rotating around the $y-$axis) and moreover it would be important to consider correlated errors from the three-qubit gate and ancilla measurement.

\section{Concluding remarks}

The numerical simulations cannot establish whether this approach would perform well in the regimes where classical approaches struggle; possibly analytical considerations may be able to rule out this promising prospect (assuming that further numerics do not already reveal some error in the data presented here). One point of concern arises from Fig. \ref{fig:timeTo999} where we see that certain 3-SAT problems are `harder' for the quantum system than others -- these harder cases evolve more slowly towards the target $\ket{\bm \theta}$ state. I do not have a complete explanation for this, although I have noted that the number of {\em nearly successful} boolean assignments is correlated to how slowly the state evolves, as one might expect. 

My original motivation for exploring these algorithms was to construct a circuit based approach to optimisation that has some inherent error tolerance, it a way analogous to adiabatic quantum computing. I am optimistic that this may indeed prove to be the case, although the preliminary simulations I have performed so far are not sophisticated enough to incorporate {\it general} noise sources, only the highly specific case of over/under rotation. But for that special case, the tolerance seems good.


\section{Acknowledgments}
My thanks to Wim van Dam, Earl Campbell and Joe Fitzsimons for helpful conversations.  

\bigskip

\bigskip

\noindent{\bf References:}
There are only a few references below -- Only those that directly crop up in the narrative. A subsequent version will add appropriate references to the literature on 3-SAT, quantum algorithms, quantum 3-SAT algorithms, and adiabatic QIP.

\bigskip
\bigskip

\noindent{\bf \large Appendix 1: Grover's algorithm}

\smallskip

Regarding the use of Grover's algorithm to directly solve a 3-SAT of 24 booleans: we have $N=2^{24}$ candidate solutions given the 3-SAT has a unique solution. We could set Grover's to perform $m$ iterates such that $\sin^2((2m+1)/\sqrt{N})$ is close to unity, which would give us $m\sim 2^{10}\pi\sim 3,217$ iterates. However because of the sine squared function, the last $\sim 20\%$ of the iterates give us a relatively poor return on the invested time. Consequently the optimal approach is ``stop early'' at $2,386$ iterations and measure; at that point the probability that our measurement gives us the correct solutions is $\sim 84\%$ but we simply restart the measurement does not give us our solution. The expected number of `runs' needed is then $1.184$. Then the expected total number of iterates to reach a solution is reduced to $2,826$, with each iterate requiring all the clauses to be tested. There are ${\rm round}(4.267*24)=102$ causes, so that the {\em expected} total number of clause checks required by Grover's algorithm is $288,252$. However it is worth reiterating that this approach requires this expected number of clause checks for {\it any} 24 boolean 3-SAT with a unique satisfying solution -- i.e. the particular features of the problem are irrelevant.

\bigskip

\noindent{\bf \large Appendix 2: Fidelity is monotonic}

\smallskip

This section considers the evolution of the qubit register's fidelity with respect to the target state, for a 3-SAT with a unique satisfying argument. However the extension to the multi-solution fidelity (next section) is straightforward. 

As observed by van Dam~\cite{vDam02}, for certain adiabatic algorithms one can construct 3-SAT problems for which the evolution `gets stuck' and never finds the actual solution. Generally this is a potential issue in any approach (quantum or classical) that seeks a solution by minimising a cost function. An example of a cost function is simply the number of unsatisfied 3-SAT clauses. An approach that involves updating the system's state so as to follow the cost function `down hill' to reach the lowest possible cost, may run the risk of being trapped in a local minimum (and therefore will require a strategy for escaping such minima). 

The present approach does not appear to have this issue: The fidelity of the qubit register with respect to the target state will only increase with each complete cycle of the clause checks. (This is not unexpected since the obvious quantum approach, i.e. Grover's algorithm, also does not involve a cost function and cannot become trapped.) Generally we can write the state of our register as 
\[
\ket{\Psi}=\alpha\ket{\bm\theta}+\beta\ket{\bm\theta^\bot}
\]
where ${\bm\theta}$ is the desired state as ${\bm\theta^\bot}$ is some state orthogonal to the desired state. The fidelity is $|\alpha|^2$ and providing that this is non-zero then when the next clause check operation is performed there is a finite probability of `passing' that check; then the post-check state is 
\[
\ket{\Psi}=\mathcal{N}(\alpha\ket{\bm\theta}+\beta^\prime\ket{\bm\theta^\bot})
\]
where $\beta^\prime\leq\beta$, normalisation constant $\mathcal{N}\geq 1$. 

The inequalities are not strict because we have not excluded the possibility that the three qubits involved were certain to pass the clause check -- they could have been in a superposition of several states, all of which satisfy the clause in question but only one of which satisfies the 3-SAT. However a {\it complete cycle} of successful clause check operations must indeed reduce $\beta$, unless it was already zero, because (as discussed earlier) $\ket{\bm\theta}$ is precisely the {\it unique} state that passes all clause checks with certainty.

Of course this leaves open the question of whether one can adversarially construct a 3-SAT problem that has an extremely slow evolution toward the target state $\ket{\bm \theta}$.
 
\bigskip

\bigskip

\noindent{\bf \large Appendix 3:}

\smallskip

\noindent{\bf \large Fidelity given multiple solutions}

\smallskip

If we have a 3-SAT with $m>1$ solutions, then applying any of the algorithms discussed here will drive the qubit register toward a subspace spanned by the states 
\begin{equation}
\ket{\bm \theta_j}=\bigotimes_i \ \ Y\left(L_{i,j}\theta\right)\ket{+}_i\ \ \ {\rm for}\ \ j=1\dots m
\label{multiSoln}
\end{equation}
where the various sets of values $\{L\}_j$ correspond to the different 3-SAT solutions in the manner specified after Eqn. (\ref{fullTheta}). Note that the value of $\theta$ is of course the same for all states; only the $L$ values vary. 

We can write a general state of our $n$ qubit register as 
\[
\ket{\Psi}=\ket{V_\parallel}+\ket{V_\bot}
\]
where $\ket{V_\parallel}$ lies in the solution subspace, and $\ket{V_\bot}$ is orthogonal to it. The probability $|\langle V_\parallel\ket{V_\parallel}|^2$ is a measure of the fidelity with which $\ket{\Psi}$ approximates an ideal solution-state, i.e. any state completely within the subspace. We may wish to record and plot this quantity during a simulation of an algorithm.

By definition
\[
\langle {\bm \theta}_j | V_\bot \rangle = 0 \ \ \ {\rm and} \ \ \ \ket{V_\parallel}=\sum_i c_i \ket{{\bm \theta}_i}
\]
remembering however that the $\ket{{\bm \theta}_j}$ are not mutually orthogonal. Then
\[
\langle {\bm \theta}_j | \Psi \rangle=\sum_i c_i \langle {\bm \theta}_j | {\bm \theta}_i \rangle\ \ \ {\rm or}\ \ \ {\bf v}={\bf M}\ {\bf c}
\]
where vector $\bf v$ has elements $\langle {\bm \theta}_j | \Psi \rangle$, vector ${\bf c}$ has elements $c_i$, and $\bf M$ is a matrix of elements $m_{i,j}= \langle {\bm \theta}_j | {\bm \theta}_i\rangle$. This matrix is easy to compute since the defition of the solution states Eqn. (\ref{multiSoln}) implies that projection $ \langle {\bm \theta}_j | {\bm \theta}_i\rangle$ is simply $(\cos\theta)^p$ where $p$ is the number of mismatching booleans (i.e. the number of instances where a given boolean is {\small TRUE} for  solution $i$ but {\small FALSE} for solution $j$, or vice versa). 

It is useful to compute the inverse matrix $\bf M^{-1}$ before beginning the simulation of an algorithm. Then at any time during the simulated execution, when the qubit register is in some state $\ket{\Psi}$, we can determine the $c_i$ by finding the elements of vector $\bf v$, i.e. the projections  $\langle {\bm \theta}_i | \Psi \rangle$, and simply evaluating
\[
{\bf c}={\bf M^{-1}}\ {\bf v}.
\]
Thus we can construct $\ket{V_\parallel}$ and moreover we can compute and record the projection on the desired subspace, $|\langle{V_\parallel}|V_\parallel\rangle|^2$.
 
\bigskip

\bigskip

\noindent{\bf \large Appendix 4:}

\smallskip

\noindent{\bf \large Information from measuring a $\theta<\frac{\pi}{2}$ state}

\smallskip

Suppose that a high fidelity approximation to the state $\ket{\bm \theta}$ with $\theta<\frac{\pi}{2}$ has been created by e.g. the sculpting algorithm. This state is simply a product of individual qubit states $Y(\pm \theta)\ket{+}$ where the sign is positive if the corresponding boolean is {\small FALSE} in the unique satisfying assignment (USA), and negative otherwise. If a given qubit is measured in the z-basis then the measurement will give the correct truth value (according to rule $\ket{0}\rightarrow$~{\small FALSE} and $\ket{1}\rightarrow$~{\small TRUE}) with a probability $p=(1+\sin\theta)/2$. In the following I will call a measurement `incorrect' if it implies the wrong boolean value; note that I do not mean to suggest some failure in the measurement apparatus (which I am taking to be perfect).  Measurements are independent of one another in this sense: the question of whether a given qubit  gives the correct value is not correlated to the correctness of any another qubit measurement. (Note that here we are considering the exact state $\ket{\bm\theta}$, whereas of course the real state created using e.g. the sculpting protocols will not be exact and may include highly correlated elements -- but for a sufficiently high fidelity the remarks here will be valid.)

Having measured all qubits and noted the results, we can repeat the process of creating $\ket{\bm \theta}$ and again measure it. This new set of measurements will be independent of the former set. So the probability of a measurement being correct is simply $p$ for every qubit and on every run.

The interesting question is, how many runs will it take until we can correctly guess the boolean truth values? We know the value of $p$ but we do not know whether a given qubit is biased towards measurement outcome $\ket{0}$ (implying the boolean is {\small FALSE}) or outcome $\ket{1}$ (implying {\small TRUE}). Assuming we always measure in the z-basis, then this scenario with an $n$ booleans is equivalent  to the following classical problem: we are given $n$ numbered coins, each with the same known bias $p>\frac{1}{2}$, but we do not know whether a given coin favours `heads' or `tails'. We are allowed to throw all coins at once and note the outcomes; how many such throws do we need before we can guess the biases correctly?

There are several ways to make this question precise and then derive an answer. I prefer the following: What is the expected number of throws, $R$, required until our {\it best guess} of the set of boolean values is {\it probability error free} -- i.e. the guess is fully correct with probability greater than a half. Now after $R$ throws we make our best guess about the bias of a given coin, i.e. head-favouring or tail-favouring, simply by noting whether it produced more heads or more tails (we can assume $R$ is odd to avoid even-splits where we don't know how to guess). Let's say a given coin is head-favouring. Then the probability that we wrongly guess `tail-favouring!' is given by the sum of the probabilities that the number of heads, $i$, is less than half of $R$.
\[
p_{\rm wrong}=\sum_{i=0}^{i=M} \left(\begin{array}{c}R \\i\end{array}\right) p^i(1-p)^{R-i},
\]
where $M=(R-1)/2$. 
In order that such a mistake will probably not occur over the entire set of $n$ coins, we require that $p_{\rm wrong}\lesssim n^{-1}$ assuming that $n$ is large, i.e. 
\begin{equation}
\sum_{i=0}^{i=M} \left(\begin{array}{c}R \\i\end{array}\right) p^i(1-p)^{R-i}<\frac{1}{n}.\nonumber
\end{equation}
\newpage
If the product $R(1-p)$ is greater than about $5$, the sum is well approximated by an integral. Making that assumption, we have:
\begin{equation}
\frac{1}{\sigma\sqrt{2\pi}}\int_{-\infty}^{R/2} \exp\left(-\frac{(x-Rp)^2}{2\sigma^2}\right)\ dx<\frac{1}{n} \nonumber
\end{equation}
where the variance $\sigma^2=Rp(1-p)$. Rescaling $x$ yields
\[
\frac{1}{\sqrt{\pi}}\int_{-\infty}^{-G} \exp(-x^2)\ dx<\frac{1}{n}\ \ \ {\rm where}\ \ \ G=\frac{\left(p-\frac{1}{2}\right)\sqrt{R}}{\sqrt{2p(1-p)}}.
\]
We recognise the complimentary error function erfc,
\[
\frac{1}{2}{\rm erfc}(G)<\frac{1}{n}
\]
and recall that to lowest order (for large $x$)
\[
{\rm erfc}(x)\simeq\frac{\exp(-x^2)}{x\sqrt{\pi}}
\]
so that the criterion becomes
\[
G\exp(G^2)>\frac{n}{2\sqrt{\pi}}.
\]
Now returning from coins to qubits and noting the assumptions made, we conclude: When attempting to solve USA 3-SAT problems of $n$ booleans using the sculpting algorithm with $R$ repetitions of the ``prepare $\ket{\bm \theta}$ and measure'' cycle, $R$ {\it need only increase logarithmically} with problem size $n$. 

Moreover we recall that bias $p=(1+\sin\theta)/2$ so we can rewrite 
\[
G=\frac{\sin\theta\sqrt{R}}{\sqrt{2(1+\sin\theta)(1-\sin\theta)}}=\sqrt\frac{R}{2}\tan\theta.
\]

We might also consider a 3-SAT of a fixed size and ask how the selected value of $\theta$ influences the expected number of repetitions $R$ needed to guess all booleans correctly. This can be assessed simply by requiring $G$ to stay constant, so 
$R\propto\cot^2\theta$ and thus $R\sim\theta^{-2}$ for small $\theta$.

\end{document}